\documentclass[12pt]{iopart}

\usepackage{graphicx}

\begin{document}

\title[Loop series for general statistical models]{Partition function loop series for a general graphical model: free energy corrections and message-passing equations}

\author{Jing-Qing Xiao$^{2,1}$ and Haijun Zhou$^{1}$}
\address{$^{1}$ Key Laboratory of Frontiers in Theoretical Physics,
Institute of Theoretical Physics, Chinese Academy of Sciences, Beijing 100190,
People's Republic of China}

\address{$^{2}$ Institute of Applied Mathematics, Academy of Mathematics and
Systems Science, Chinese Academy of Sciences, Beijing 100190,
People's Republic of China}

\ead{zhouhj@itp.ac.cn}

\begin{abstract}
A loop series expansion for the partition function of a general statistical
model on a graph is carried out.
If the auxiliary probability distributions of the
expansion are chosen to be a fixed point of the belief-propagation equation,
the first term of the loop series  gives the
Bethe-Peierls free energy functional at the replica-symmetric
level of the mean-field
spin glass theory, and corrections are contributed only by subgraphs
that are free of dangling edges. This result generalize the early work of
Chertkov and Chernyak on binary statistical models.
If the belief-propagation
equation has multiple fixed points, a loop series expansion
is performed for the grand partition function.
The first term of this series gives the Bethe-Peierls free energy functional at
the first-step replica-symmetry-breaking (RSB) level of the mean-field spin-glass theory,
and corrections again come only from subgraphs that are free of dangling edges,
provided that the
auxiliary probability distributions of the expansion are chosen to be a
fixed point of the survey-propagation equation.
The same loop series expansion
can be performed for higher-level partition functions, obtaining
the higher-level RSB Bethe-Peierls free energy functionals (and the correction terms)
and message-passing equations without using the Bethe-Peierls approximation.
\end{abstract}


\vspace{2pc}

\submitto{\JPA}
\maketitle

\section{Introduction}
Graph expansion methods for statistical models have been extensively
discussed in the literature. They have been very helpful in studying the
high-temperature behaviours and
the phase transition properties of discrete models such as the Ising model of
ferromagnetism.
Some of the early efforts were carried out by Brout and others
\cite{Brout-1959,Horwitz-Callan-1961,Englert-1963}.
More recently, Georges and Yedidia \cite{Georges-Yedidia-1991} found
that, high-temperature expansion of the Ising spin glass free energy
can also be carried out under the constraints
of fixed mean spin values. This later work was extended
by Sessak and Monasson \cite{Sessak-Monasson-2009} to include
pair correlations  of spin variables as another set of expansion constraints.
The constrained graph expansion method was applied to
the inverse Ising problem \cite{Sessak-Monasson-2009},
with the aim of  inferring the microscopic interactions of a Ising
system based on the observed mean spin values and spin-pair correlations.

For finite-connectivity binary statistical models, Chertkov and Chernyak
\cite{Chertkov-Chernyak-2006a,Chertkov-Chernyak-2006b} showed that the
partition function can be expressed as a sum of contributions from
subgraphs.
The first term of this expansion is identical to the partition
function obtained using the Bethe-Peierls (BP) approximation. Loop corrections to
the BP approximation was also calculated by Montanari, Rizzo, and
collaborators \cite{Montanari-Rizzo-2005,Rizzo-Wemmenhove-Kappen-2007} and by
Parisi and Slanina \cite{Parisi-Slanina-2006}.
The derivation of the partition function expansion
by Chertkov and Chernyak relied on a special property of Ising spin variables
(see equation (17) of \cite{Chertkov-Chernyak-2006b}). This special property
is not valid for more general statistical models, whose microscopic variables
are not necessarily binary or discrete.
Whether the conclusion of
\cite{Chertkov-Chernyak-2006a,Chertkov-Chernyak-2006b} is valid to
general statistical models is still an open issue
(for models whose
discrete variables take $Q>2$ values, a complicated
loop-tower expansion was presented in \cite{Chernyak-Chertkov-2007}).

In the present contribution we first extend the results of
\cite{Chertkov-Chernyak-2006a,Chertkov-Chernyak-2006b,Chernyak-Chertkov-2007}
by carrying out a very simple derivation
of partition function loop series for a general statistical model
defined on a graph. We do not make any assumptions on the
nature of the microscopic state variable of each
edge (or vertex) of the graph. This state variable can be discrete or
real-valued, or be a vector, or even be a function itself.
We show  that the first term of this expansion
is also identical to the Bethe-Peierls (BP) partition function,
and corrections to the BP partition function come only from looped
subgraphs without any dangling edges. The auxiliary probability distributions
of this loop series expansion are chosen to be a fixed point of the
belief-propagation equation. This particular choice makes all the
subgraphs with at least one dangling edge to have zero contribution to
the correction terms.

As the second main result, we present the loop series expressions for
the grand partition function and higher-level
partition functions. The belief-propagation equation of a statistical model may have multiple
fixed points, each of which  is referred to as a macroscopic state of the
configuration space. If this happens, we define a grand partition
function at the level of macroscopic states and perform a loop series
expansion for this grand partition function. When the auxiliary
probability distributions of this expansion are chosen to be a fixed
point of the survey-propagation equation (first derived through the first-step
replica-symmetry-breaking (1RSB) mean-field theory of  spin glasses \cite{Mezard-Parisi-2001}), the
first term of this loop expansion gives the BP free energy functional at the level of
macroscopic states. Corrections again
come only from subgraphs that are free of dangling edges. In case the survey-propagation
equation has multiple fixed points, the same loop
series expansion can be performed for higher-level partition functions. As a result,
we obtain the higher-level BP free energy functionals and the correction contributions, and
the associated message-passing equations.

This work is a mathematical approach to the theory of spin glasses
from the framework of partition function loop
expansion. It is clear that at each replica-symmetry-breaking (RSB) level of the
mean-field theory \cite{Mezard-Parisi-2001} the corrections to the
free energy due to looped nontrivial subgraphs are neglected.
This neglected correction contribution is explicitly expressed as a logarithm over a
finite series in this paper. At a given level of macroscopic states we anticipate that,
the magnitude of the total loop correction
contribution to the free energy will be sub-linear in $N$
($N$ being the total number of vertices) if there is only one fixed-point
for the corresponding message-passing equation, but it will be
linear in $N$ if there exist multiple fixed-points.
This statement needs to be checked by numerical calculations on single graphical systems.

Section~\ref{sec:model} introduces the general statistical model.
We work out the loop series of the partition function in Sec.~\ref{sec:loop} and
derive the belief-propagation equation.
In Sec.~\ref{sec:rsb} we extend the discuss to the case that the
belief-propagation equation has multiple fixed points, and perform a loop
series expansion for the grand partition function. The 1RSB survey-propagation
equation is derived here.
We conclude this work in Sec.~\ref{sec:conclude}, and discuss some possible
extensions. The \ref{sec:ring} contains graph expansion
results for a one-dimensional ring.

\section{General statistical models on graphs}
\label{sec:model}

We consider a graph $G$ composed of $N$ vertices ($i=1,2,\ldots, N$) and
$M$ edges. An edge $(i,j)$ between two vertices $i$
and $j$ has a state variable $x_{i j}$.
This variable may be a binary spin for some systems, $x_{i j}=\pm 1$. For other systems,
$x_{i j}$ may be real-valued or be a vector, or be even more complicated.
In this paper we make no assumptions
on the nature of the microscopic state variable $x_{i j}$ of each edge $(i,j)$.
Each vertex $i$ has an energy $E_i(x_{i \partial i})$, where
$x_{i \partial i}\equiv \{x_{i j_1}, x_{i j_2}, \ldots,
x_{i j_k}\}$ with $j_1, j_2,\ldots, j_k$ being the $k$
other vertices with which $i$ forms an edge.
The number $k$ of nearest neighbors
of a vertex might be different for different vertices.
The vertex energy is a function of the
state variables of its connected edges.
Notice that $x_{i j}$ and $x_{j i}$ both denote
the state of edge $(i,j)$, therefore $x_{i j}\equiv x_{j i}$.
The total vertex energy for an edge configuration $\{x_{i j}\}$ is
\begin{equation}
E(\{x_{i j}\})=\sum\limits_{i=1}^{N} E_{i}(x_{i \partial i}) .
\end{equation}
The partition function of the system is defined as
\begin{equation}
\label{eq:partitionf}
Z(\beta) = \prod\limits_{(i,j)}
\Bigl[\int {\rm d} x_{i j} \rho_0(x_{i j})\Bigr]
 \prod\limits_{i=1}^{N} e^{-\beta E_i(x_{i \partial i})} .
\end{equation}
In the above equation, $\beta$ is the inverse temperature,
$\rho_0(x_{i j})$ is the probability of
microscopic state $x_{i j}$ for an isolated edge $(i, j)$,
and $\prod_{(i,j)}$ means the product over all the edges of
graph $G$. For simplicity we
assume that the {\em a priori} probabilities $\rho_0(x_{i j})$ are identical
for all edges. This assumption is of cause nonessential.

The partition function (\ref{eq:partitionf}) also applies to graphical
models whose microscopic states are defined on vertices rather than on
edges \cite{Chertkov-Chernyak-2006a,Chertkov-Chernyak-2006b}.
For example, consider a graph $G$ with the property that its vertices can be
divided into two subsets, denoted by $\{i\}$ (the variable nodes) and
$\{a\}$ (the check nodes), such that all the edges of $G$ are between
a variable node $i$ and a check node $a$.
For each variable node $i$ we assume that
\begin{equation}
e^{-\beta E_i(x_{i \partial i})}=
\int {\rm d} x_i \rho_0(x_i)
\prod\limits_{a\in \partial i}
 \Bigl[\frac{\delta(x_{i a}-x_i)}{\rho_0(x_{i a})}\Bigr] ,
\end{equation}
where $\partial i$ denotes the set of nearest neighboring check nodes of
$i$. \Eref{eq:partitionf} then
simplifies to
\begin{equation}
\label{eq:20110417-01}
 Z(\beta) = \prod\limits_{i}
\Bigl[\int {\rm d} x_{i} \rho_0(x_{i})\Bigr]
 \prod\limits_{a} e^{-\beta E_a(x_{\partial a})} ,
\end{equation}
where $\partial a$ denotes the set of nearest neighboring variable
nodes of $a$.
Equation (\ref{eq:20110417-01}) is the partition function of a system
defined on a  factor graph, with each variable node $i$ having
a microscopic state $x_i$ and each check node $a$ having an energy
$E_a$. The check energy $E_a$ depends on the microscopic state
$x_{\partial a}$ of the variable nodes in $\partial a$.

In some graphical models, the state $x_{i j}$ of each edge $(i,j)$
is a collection of two microscopic states $y_{i j}^i$ and $y_{i j}^j$,
$x_{i j}\equiv \{y_{i j}^i, y_{i j}^j\}$.
We assume that the
{\em a priori} probability distribution of the edge state $x_{i j}$ equals to
$\rho_0(y_{i j}^i) \rho_0(y_{i j}^j)$, and that
the energy $E_i$ of a vertex $i$ can be expressed as
\begin{equation}
 e^{-\beta E_i(x_{i \partial i})}=
\int {\rm d} y_{i} \rho_0(y_i)
e^{-\beta \tilde{E}_i(y_i, \{y_{i j}^j\})}
\prod\limits_{j\in \partial i}
\Bigl[
\frac{\delta (y_i-y_{i j}^{i})}
{\rho_0(y_{i j}^{j})}
\Bigr] .
\end{equation}
Under these assumptions, the partition function (\ref{eq:partitionf}) becomes
\begin{equation}
 Z(\beta) = \prod\limits_{i}
\Bigl[\int {\rm d} y_{i} \rho_0(y_{i})
 e^{-\beta \tilde{E}_i(y_i, y_{\partial i})} \Bigr] ,
\end{equation}
which describes a graphical model whose vertex energy
$\tilde{E}_i$ depends on the microscopic state $y_i$ of
vertex $i$ and the microscopic states $y_{\partial i}$ of
its nearest neighbors.
An example of such statistical models is the palette-coloring problem
\cite{Bounkong-vanMourik-Saad-2006,Wong-Saad-2008,Pelizzola-Pretti-vanMourik-2011}.

\section{Graph expansion for the general statistical model}
\label{sec:loop}

To find a loop series expression for the partition function
(\ref{eq:partitionf}), we introduce for each edge $(i,j)$ two
auxiliary probability distributions $q_{j\rightarrow i}(x_{i j})$ and
$q_{i\rightarrow j}(x_{j i})$, and rewrite $Z(\beta)$ as
\begin{eqnarray}
\fl
\quad\quad\quad
Z(\beta) &=& \prod\limits_{i=1}^{N} \prod\limits_{j\in \partial i}
\Bigl[ \int {\rm d} x_{i j} q_{j\rightarrow i}(x_{i j}) \Bigr]
e^{-\beta E_i(x_{i \partial i})} \prod\limits_{(k,l)}
\frac{\delta(x_{k l}-x_{l k}) \rho_0(x_{k l})}{
q_{k\rightarrow l}(x_{l k}) q_{l\rightarrow k}(x_{k l})}
\label{eq:pf2} \\
\fl
&=& \frac{1}{\prod\limits_{(i,j)} C_{(i, j)}}
\prod\limits_{i=1}^{N}
\prod\limits_{j\in \partial i}
\Bigl[ \int {\rm d} x_{i j} q_{j\rightarrow i}(x_{i j})\Bigr]
e^{-\beta E_i(x_{i \partial i})}
 \prod\limits_{(k,l)}\Bigl[1+ \Delta_{(k, l)}(x_{k l}, x_{l k})\Bigr] .
\label{eq:20110410-02}
\end{eqnarray}
In (\ref{eq:20110410-02}), $C_{(i, j)}$ is an edge constant with the value
\begin{equation}
C_{(i, j)} = \int {\rm d} x_{i j} \frac{q_{i\rightarrow j}(x_{i j})
q_{j\rightarrow i} (x_{i j})}{\rho_0(x_{i j})} ,
\end{equation}
and $\Delta_{(i, j)}(x_{i j}, x_{j i})$ is expressed as
\begin{equation}
\Delta_{(i, j)}(x_{i j}, x_{j i}) \equiv
\frac{\delta(x_{i j}-x_{j i}) \rho_0(x_{i j}) C_{(i, j)}}{
q_{i\rightarrow j}(x_{j i}) q_{j\rightarrow i}(x_{i j})} - 1 .
\end{equation}
From (\ref{eq:20110410-02}) we know that the partition
function can be expressed as the sum of contributions
from all the possible non-empty subgraphs of $G$:
\begin{equation}
\label{eq:20110410-03}
Z(\beta)=Z_{BP}\Bigl(1 + \sum\limits_{g \subseteq G} L_{g}\Bigr) .
\end{equation}
In the above equation, $Z_{BP}$ is calculated by
\begin{equation}
Z_{BP} = \frac{\prod_{i=1}^{N} \prod_{j\in \partial i}
\Bigl[ \int {\rm d} x_{i j} q_{j\rightarrow i}(x_{i j}) \Bigr]
e^{-\beta E_i(x_{i \partial i})}}
{\prod_{(i,j)}\Bigl[ \int {\rm d} x_{i j}
 \frac{q_{i\rightarrow j}(x_{i j})
q_{j\rightarrow i}(x_{i j})}{\rho_0(x_{i j})}\Bigr]} .
\end{equation}
A non-empty subgraph $g$ of graph $G$ contains a
partial set of the edges of $G$ and all the vertices that are attached
to these edges. The correction $L_g$ is expressed as
\begin{equation}
\label{eq:20110801-01}
L_g = \prod\limits_{i\in g}
\frac{ \prod_{j\in \partial i} \Bigl[\int
{\rm d} x_{i j} q_{j\rightarrow i}(x_{i j}) \Bigr]
e^{-\beta E_i(x_{i \partial i})}
}{\prod_{j\in \partial i} \Bigl[\int
{\rm d} y_{i j} q_{j\rightarrow i}(y_{i j}) \Bigr]
e^{-\beta E_i(y_{i \partial i})} }
\prod\limits_{(k,l) \in g} \Delta_{(k, l)}(x_{k l}, x_{l k}) .
\end{equation}

\begin{figure}
\begin{center}
\includegraphics[width=0.4\textwidth]{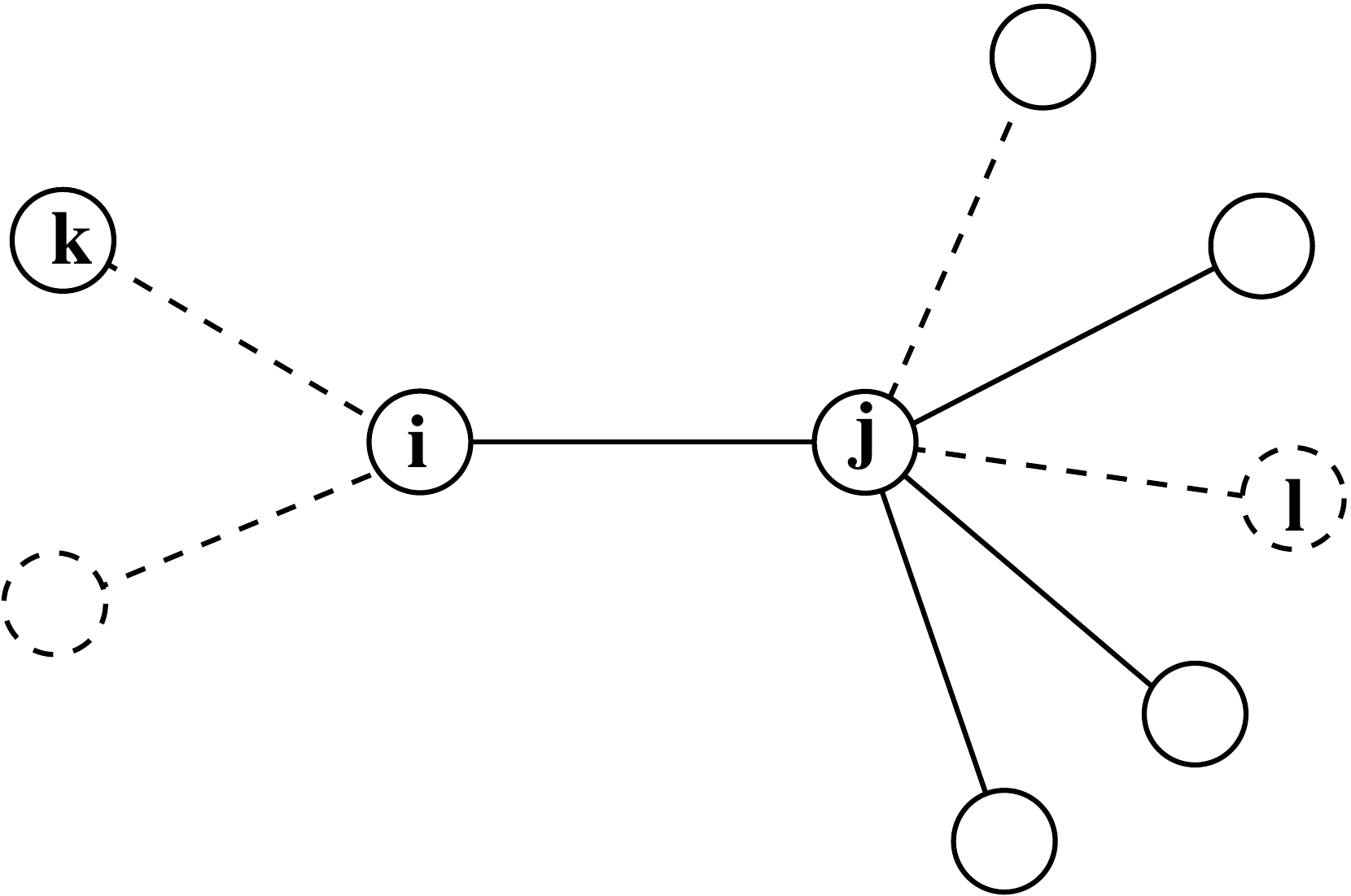}
\end{center}
\caption{\label{fig:20110411-01}
The neighborhood of a vertex $i$. A solid line between any two vertices
$i$ and $j$ means that the edge $(i,j)$ is
presented both in graph $G$ and in subgraph $g$. A dashed line between
two vertices $i$ and $k$ means that the edge
$(i,k)$ is presented in $G$ but not in $g$.
A solid circle denotes a vertex that belongs to subgraph $g$,
and a dashed circle denotes a vertex that is
not belonging to $g$. A solid circle is attached by at least
one solid edge. In this figure, vertex $i$ is connected by only one solid edge, it is
a leaf vertex of subgraph $g$, and the edge $(i,j)$ is a dangling edge.
}
\end{figure}

Consider a subgraph $g$ which has a vertex $i$ that is linked to the other
parts of $g$ only through a single edge $(i,j)$. The neighborhood of such
a leaf vertex $i$ is shown schematically in
figure~\ref{fig:20110411-01}. We find that after integrating
over the variable $x_{i j}$, the correction $L_g$ is expressed as
\begin{eqnarray}
\fl
\quad\quad\quad\quad
L_g & =&  \prod\limits_{k \in g\backslash i}
\frac{ \prod_{l\in \partial k} \Bigl[\int
 {\rm d} x_{k l} q_{l\rightarrow k}(x_{k l})\Bigr]
e^{-\beta E_k(x_{k \partial k})}
}{ \prod_{l\in \partial k} \Bigl[\int
{\rm d} y_{k l} q_{l\rightarrow k}(y_{k l}) \Bigr]
e^{-\beta E_k(y_{k \partial k})}
}
\prod\limits_{(m,n) \in g\backslash (i,j)}
\Delta_{(m,n)}(x_{m n}, x_{n m}) \nonumber \\
& & \quad\quad \times
\left\{
\frac{\hat{q}_{i\rightarrow j}(x_{j i})
 \int {\rm d} y_{i j} q_{j\rightarrow i}(y_{i j})
q_{i\rightarrow j}(y_{i j})/\rho_0(y_{i j})}
{q_{i\rightarrow j}(x_{j i})
\int {\rm d} y_{i j} q_{j\rightarrow i}(y_{i j})
 \hat{q}_{i\rightarrow j}(y_{i j})/\rho_0(y_{i j})} -1
\right\} ,  \label{eq:20110411-05}
\end{eqnarray}
where $\hat{q}_{i\rightarrow j}(x_{i j})$ is determined by
the set of probability functions $q_{\partial i\backslash j}\equiv
\{q_{k\rightarrow i}, k\in \partial i\backslash j\}$ through
\begin{equation}
\label{eq:20110417-02}
\fl
\quad\quad
\hat{q}_{i\rightarrow j}(x_{i j}) =
B_{i\rightarrow j}(q_{\partial i\backslash j}) \equiv \frac{\rho_0(x_{i j})
\prod_{k\in \partial i\backslash j}
\Bigl[\int
{\rm d} x_{i k} q_{k\rightarrow i}(x_{i k})
\Bigr] e^{-\beta E_i(x_{i \partial i})} }{
\int {\rm d} y_{i j} \rho_0(y_{i j})
\prod_{k\in \partial i\backslash j}
\Bigl[ \int  {\rm d} y_{i k}
q_{k\rightarrow i}(y_{i k})
\Bigr] e^{-\beta E_i(y_{i \partial i})}} .
\end{equation}
The function $B_{i\rightarrow j}(q_{\partial i\backslash j})$ as defined by (\ref{eq:20110417-02}) is called the belief-propagation equation. It takes as
input a set of probability distributions
$q_{k\rightarrow i}(x_{i k})$ ($k\in \partial i$) and outputs a
new probability distribution $\hat{q}_{i\rightarrow j}(x_{i j})$.

Since we are free to choose the auxiliary
probabilities functions $\{q_{i\rightarrow j}(x_{j i})\}$, we
can choose this set of auxiliary functions to be a fixed point of the
belief-propagation equation (\ref{eq:20110417-02}). In other words, we
require that the auxiliary probability functions to satisfy
\begin{equation}
\label{eq:BPfix}
q_{i\rightarrow j}(x_{i j}) = B_{i\rightarrow j}\Bigl(
\{q_{k\rightarrow i}(x_{i k}), k\in \partial i\backslash j\}\Bigr) .
\end{equation}
Then for each edge $(i,j)$ we have $\hat{q}_{i\rightarrow j}(x_{i j}) =
{q}_{i\rightarrow j}(x_{i j})$, and
the expression inside the curly brackets of (\ref{eq:20110411-05}) is
identically zero.  Under this special choice, only those subgraphs of graph $G$
with each vertex $i$ having at least two attached edges have non-zero
contributions to the correction of the partition function.
The total free energy $F(\beta)$ is then expressed as
\begin{equation}
\label{eq:ffull}
F(\beta) \equiv -\frac{1}{\beta} \ln Z(\beta) =
F_{BP}(\beta) -\frac{1}{\beta} \ln\Bigl[
1+ \sum\limits_{{g^\prime}\subseteq G} L_{{g^\prime}}\Bigr],
\end{equation}
where ${g^\prime}$ denotes a looped subgraph that contains no dangling edges. The
free energy $F_{BP}(\beta)$ corresponds to the partition function $Z_{BP}$
and is expressed as
\begin{equation}
\label{eq:20110415-01}
F_{BP}(\beta) = \sum\limits_{i} f_i - \sum\limits_{(i,j)} f_{(i,j)} ,
\end{equation}
with
\begin{eqnarray}
f_i &=& -\frac{1}{\beta} \ln\left[
\prod\limits_{k\in \partial i} \Bigl[
\int {\rm d} x_{i k} q_{k\rightarrow i}(x_{i k}) \Bigr]
e^{-\beta E_i(x_{i \partial i})} \right] , \label{eq:20110417-03} \\
f_{(i,j)} &=& -\frac{1}{\beta} \ln\left[
\int {\rm d} x_{i j} \frac{q_{i\rightarrow j}(x_{i j})
	 q_{j\rightarrow i}(x_{i j})}{\rho_0(x_{i j})}
\right] . \label{eq:20110417-04}
\end{eqnarray}

We emphasize that $F_{BP}(\beta)$ is identical in form to the mean-field
free energy as obtained by the replica-symmetric (RS) spin-glass theory
\cite{Mezard-Parisi-2001}. The expression (\ref{eq:20110415-01}) was
first derived in the mean-field theory by using the BP approximation.
The free energy $F_{BP}$ can also be viewed as a functional of the $2 M$ probability
distributions $\{p_{i\rightarrow j}(x_{i j})\}$ on the $M$ edges $(i,j)$ of graph
$G$. In this paper we refer $F_{BP}$ as the BP free energy functional. It is
easy to check that the variation of $F_{BP}$ with respect to
any a probability distribution $p_{i\rightarrow j}(x_{i j})$ is zero at a fixed point of
 (\ref{eq:BPfix}).
\Eref{eq:20110415-01} is expressed as the difference between the total vertex contribution ($\sum_i f_i$) and the total edge contribution ($\sum_{(i,j)} f_{(i,j)}$). An intuitive understanding of this is as follows: Each edge participates in two vertex interactions and its effect is counted twice when calculating the total vertex contribution; this over-counting should be subtracted from the total vertex contribution.

From the viewpoint of partition function loop expansion, the
belief-propagation fixed-point condition \eref{eq:BPfix} is a requirement
for ensuring all the corrections $L_g$ from subgraphs $g$ with dangling
edges are identically zero. For a loopy subgraph $g$ without dangling edges, its correction contribution $L_g$ is obtained through \eref{eq:20110801-01}. The correction
to the total free energy is expressed as the logarithm of the sum of all
these loop correction contributions [see (\ref{eq:ffull})]. In \ref{sec:ring} we report
the free energy correction contribution of a one-dimensional ring of $N$ edges. The correction is found to be positive when this ring is energetically
frustrated. For more complicated model systems, the sign of the free energy correction contribution is still an open issue.

For a discrete model whose edge states can take  $Q>2$ different values,
Chernyak and Chertkov \cite{Chernyak-Chertkov-2007} derived a loop-tower expansion for the partition function by exploiting the
gauge symmetry of the microscopic states. The derived belief-propagation equation by their approach does not
fix the gauge freedom completely, and therefore high-order gauge fixing was introduced, making the loop-tower expansion
scheme very complicated. Gauge fixing is not needed in the present loop series expansion scheme.
In light of the present work, it might be possible to simplify the scheme of \cite{Chernyak-Chertkov-2007} and
get an alternative derivation of
the free energy expression (\ref{eq:ffull}). We are currently working on this mathematical issue.

\section{Graph expansion for the grand partition function}
\label{sec:rsb}

For the general statistical model defined by the partition
function (\ref{eq:partitionf}), the belief-propagation equation
(\ref{eq:BPfix}) may have multiple fixed points. If this happens, then
the BP
free energy $F_{BP}$ as a functional of $\{p_{i\rightarrow j}(x_{i j})\}$
has more than one minimal value. In the following, we will refer to a
fixed-point solution $\{p_{i\rightarrow j}(x_{i j})\}$ of
(\ref{eq:BPfix}) with a minimal value of $F_{BP}$ as a macroscopic state of
the configuration space. Each macroscopic state has a corresponding BP
free energy value $F_{BP}$.
To account for the existence of multiple macroscopic states, in analogy with
(\ref{eq:partitionf}), we
define a grand partition function $\Xi$ as
\begin{eqnarray}
\fl
 \Xi = \prod\limits_{(i,j)} \Bigl[
\int \int {\rm D} q_{i\rightarrow j}
{\rm D} q_{j\rightarrow i}
\delta\Bigl(q_{i\rightarrow j}-B_{i\rightarrow j}(q_{\partial i\backslash j})
\Bigr) \delta\Bigl(q_{j\rightarrow i}-B_{j\rightarrow i}(q_{\partial j
\backslash i})\Bigr) \Bigr] \exp(-y F_{BP}) .
\label{eq:20110415-02}
\end{eqnarray}
In the above equation, $\int {\rm D} q_{i\rightarrow j}$ means
summing over all different possibilities of the distribution
$q_{i\rightarrow j}$, and the Dirac delta functions
$\delta\Bigl(q_{i\rightarrow j}-B_{i\rightarrow j}
(q_{\partial i\backslash j})\Bigr)$
ensure that only fixed-point solutions of the belief-propagation equation
(\ref{eq:BPfix}) contribute to $\Xi$. The parameter $y$ is an introduced
inverse temperature at the level of macroscopic states.

In analogy with (\ref{eq:pf2}) we can rewrite (\ref{eq:20110415-02}) as
\begin{equation}
\Xi = \frac{1}{\prod_{(i,j)} C_{(i, j)}^{(1)}}
\prod\limits_{i=1}^{N} \Bigl[ \prod\limits_{j\in \partial i}
\int {\rm D} q_{j\rightarrow i}
P_{j\rightarrow i}(q_{j\rightarrow i})
e^{-y f_i} \Bigr]
\prod\limits_{(k,l)} \Bigl[1+ \Delta_{(k, l)}^{(1)}\Bigr] .
\end{equation}
In the above equation, $P_{i\rightarrow j}(q_{i\rightarrow j})$ is
an introduced auxiliary probability distribution function for the probability
distribution $q_{i\rightarrow j}(x_{j i})$; $f_i$
is the free energy contribution of vertex $i$ as expressed by
(\ref{eq:20110417-03}); $C_{(i,j)}^{(1)}$ is an edge constant,
\begin{equation}
C_{(i, j)}^{(1)}= \int \int {\rm D} q_{i\rightarrow j}
{\rm D} q_{j\rightarrow i}
P_{i\rightarrow j}(q_{i\rightarrow j})
P_{j\rightarrow i}(q_{j\rightarrow i}) e^{-y f_{(i,j)}} ,
\end{equation}
with $f_{(i,j)}$ being the free energy contribution of an edge $(i,j)$ as given
by  (\ref{eq:20110417-04}); and $\Delta_{(i,j)}^{(1)}$ is expressed as
\begin{equation}
 \Delta_{(i,j)}^{(1)} \equiv
\frac{\delta\Bigl(q_{i\rightarrow j}-B(q_{\partial i\backslash j})\Bigr)
\delta\Bigl(q_{j\rightarrow i}-B(q_{\partial j\backslash i})\Bigr)
{C}_{(i, j)}^{(1)}}
{P_{i\rightarrow j}(q_{i\rightarrow j})
 P_{j\rightarrow i}(q_{j\rightarrow i}) e^{-y f_{(i,j)}}}-1 .
\end{equation}
The grand partition function can therefore be expanded as
\begin{equation}
\label{eq:20110416-01}
\Xi= \Xi_{SP} \Bigl(1+ \sum\limits_{g\subseteq G} L_g^{(1)}\Bigr) ,
\end{equation}
where $\Xi_{SP}$ is expressed as
\begin{equation}
 \Xi_{SP}=\frac{
\prod_{i=1}^{N} \Bigl[
\prod_{j\in \partial i} \int {\rm D} q_{j\rightarrow i}
P_{j\rightarrow i}(q_{j\rightarrow i}) e^{-y f_i} \Bigr]
}
{\prod_{(i,j)} \Bigl[
\int \int {\rm D} q_{i\rightarrow j} {\rm D} q_{j\rightarrow i}
P_{i\rightarrow j}(q_{i\rightarrow j})
P_{j\rightarrow i}(q_{j\rightarrow i})
e^{-y f_{(i, j)}} \Bigr]
} ,
\end{equation}
and  the correction term $L_g^{(1)}$ of a subgraph $g$
is expressed as
\begin{equation}
 L_g^{(1)} =
\prod\limits_{i\in g}
\frac{
\prod_{j\in \partial i} \Bigl[\int
{\rm D} q_{j\rightarrow i} P_{j\rightarrow i}(q_{j\rightarrow i})
\Bigr] e^{-\beta f_i(\{q_{j\rightarrow i}\})} }{
\prod_{j\in \partial i}
\Bigl[\int {\rm D} p_{j\rightarrow i} P_{j\rightarrow i}
(p_{j\rightarrow i}) \Bigr]
 e^{-\beta f_i(\{p_{j\rightarrow i}\})} } \prod\limits_{(k,l) \in g} \Delta_{(k,l)}^{(1)} .
\end{equation}

Consider a subgraph $g$ which has a leaf vertex $i$ and a dangling edge
$(i,j)$. After integrating over the probabilities around vertex $i$, the
correction contribution of this subgraph can be expressed as
\begin{eqnarray}
\fl
\quad\quad
L_g^{(1)} =  \prod\limits_{k\in g \backslash i}
\frac{\prod_{l\in \partial k} \Bigl[\int
{\rm D} q_{l\rightarrow k} P_{l\rightarrow k}(q_{l\rightarrow k})
\Bigr] e^{-\beta f_k(\{q_{l\rightarrow k}\})} }{\prod_{l\in \partial k}
\Bigl[\int {\rm D} p_{l\rightarrow k} P_{l\rightarrow k}
(p_{l\rightarrow k}) \Bigr]
 e^{-\beta f_k(\{p_{l\rightarrow k})\}} } \prod\limits_{(m,n) \in g \backslash (i,j)}
 \Delta_{(m,n)}^{(1)} \nonumber \\
\times \left\{
\frac{\hat{P}_{i\rightarrow j}(q_{i\rightarrow j}) \int \int {\rm D}
q_{i\rightarrow j} {\rm D} q_{j\rightarrow i}
P_{i\rightarrow j}(q_{i\rightarrow j}) P_{j\rightarrow i}(q_{j\rightarrow i})
 e^{-y f_{(i,j)}}
}{
P_{i\rightarrow j}(q_{i\rightarrow j}) \int \int {\rm D} q_{i\rightarrow j}
{\rm D} q_{j\rightarrow i} \hat{P}_{i\rightarrow j}(q_{i\rightarrow j})
P_{j\rightarrow i}(q_{j\rightarrow i}) e^{-y f_{(i,j)}}
}-1
\right\} ,
\label{eq:20110419-01}
\end{eqnarray}
where the probability distribution
$\hat{P}_{i\rightarrow j}(q_{i\rightarrow j})$ is
calculated by
\begin{equation}
\label{eq:surveyprop}
\fl
\quad\quad\quad\quad
\hat{P}_{i\rightarrow j}(q_{i\rightarrow j})
=\frac{
\prod_{k\in \partial i\backslash j} \Bigl[ \int
{\rm D} q_{k\rightarrow i} P_{k\rightarrow i}(q_{k\rightarrow i})
\Bigr] e^{-y f_{i\rightarrow j}}
 \delta\Bigl(q_{i\rightarrow j}-B_{i\rightarrow j}
(q_{\partial i\backslash j})\Bigr)
}{
\prod_{k\in \partial i\backslash j} \Bigl[ \int
{\rm D} q_{k\rightarrow i} P_{k\rightarrow i}(q_{k\rightarrow i})
\Bigr] e^{-y f_{i\rightarrow j}} } ,
\end{equation}
with
\begin{equation}
 f_{i\rightarrow j}=-\frac{1}{\beta}\ln\Bigl[
\int {\rm d} x_{i j} \rho_0 (x_{i j})
\prod\limits_{k\in \partial i\backslash j} \int {\rm d} x_{i k}
q_{k\rightarrow i}(x_{i k}) e^{-\beta E_i (x_{i \partial i})}
\Bigr] .
\end{equation}
In accordance with the spin-glass literature, we refer (\ref{eq:surveyprop}) as
the the survey-propagation equation.

The expression inside the curly brackets of (\ref{eq:20110419-01}) is
identically zero if $\hat{P}_{i\rightarrow j}(q_{i\rightarrow j})
= P_{i\rightarrow j}(q_{i\rightarrow j})$.
Since the auxiliary probability distributions
$\{P_{i\rightarrow j}(q_{i\rightarrow j})\}$
are free to choose, we can choose them appropriately to
ensure that the correction contribution $L_g^{(1)}=0$ for any a
subgraph $g$ with at least one dangling edge.
In other words, $\{P_{i\rightarrow j}(q_{i\rightarrow j})\}$ should be
a fixed-point solution of the survey-propagation equation:
\begin{equation}
\label{eq:SPfix}
\fl
\quad\quad\quad\quad
 P_{i\rightarrow j}(q_{i\rightarrow j})
=\frac{
\prod_{k\in \partial i\backslash j} \Bigl[ \int
{\rm D} q_{k\rightarrow i} P_{k\rightarrow i}(q_{k\rightarrow i})
\Bigr] e^{-y f_{i\rightarrow j}}
 \delta\Bigl(q_{i\rightarrow j}-
B_{i\rightarrow j}(q_{\partial i\backslash j})\Bigr)
}{
\prod_{k\in \partial i\backslash j} \Bigl[ \int
{\rm D} q_{k\rightarrow i} P_{k\rightarrow i}(q_{k\rightarrow i})
\Bigr] e^{-y f_{i\rightarrow j}} } .
\end{equation}
This equation was first derived in \cite{Mezard-Parisi-2001} under physical considerations (the BP
approximation was again used).

At a fixed point of (\ref{eq:SPfix}), the expression of the total grand free energy is
\begin{equation}
\label{eq:gfe}
G(y; \beta) \equiv -\frac{1}{y} \ln \Xi
= G_{SP}(y; \beta) - \frac{1}{y} \ln \Bigl[
1+\sum\limits_{{g^\prime}\subseteq G} L_{{g^\prime}}^{(1)}\Bigr] ,
\end{equation}
where ${g^\prime}$ again denotes a looped subgraph that contains no
dangling edges. From the framework of partition function loop expansion,
(\ref{eq:SPfix}) is a requirement to ensure that subgraphs with
dangling edges do not have correction contributions to the grand free energy.

In (\ref{eq:gfe}), the grand free energy $G_{SP}(y;\beta)$ is expressed as
\begin{equation}
\label{eq:20110416-03}
 G_{SP}(y;\beta) \equiv -\frac{1}{y}\ln \Xi_{SP}
 = \sum\limits_{i} g_i - \sum\limits_{(i,j)}
g_{(i,j)} ,
\end{equation}
with
\begin{eqnarray}
 g_i &=& -\frac{1}{y} \ln\Bigl[
\prod_{j\in \partial i} \int {\rm D} q_{j\rightarrow i}
P_{j\rightarrow i}(q_{j\rightarrow i}) e^{-y f_i}
\Bigr] ,\\
g_{(i,j)} &=&-\frac{1}{y}\ln\Bigl[
\int \int {\rm D} q_{i\rightarrow j} {\rm D} q_{j\rightarrow i}
P_{i\rightarrow j}(q_{i\rightarrow j})
P_{j\rightarrow i}(q_{j\rightarrow i})
e^{-y f_{(i, j)}}
\Bigr]
\end{eqnarray}
being, respectively, the contribution to the grand free energy from a
vertex $i$ and an edge $(i,j)$. $G_{SP}(y; \beta)$
is identical in form to the 1RSB free energy of the mean-field
spin-glass theory \cite{Mezard-Parisi-2001}, which was derived previously
by applying the BP approximation. $G_{SP}$ can also be
regarded as a functional of the $2 M$ probabilities
$\{P_{i\rightarrow j}(q_{i\rightarrow j})\}$, and its variation with respect to
any a $P_{i\rightarrow j}(q_{i\rightarrow j})$ is zero at a fixed-point of
the survey-propagation equation. We refer $G_{SP}$ as the survey-propagation free
energy functional (it is the BP free energy functional at the 1RSB mean-field
level).

We end this section with a discussion on the reweighting parameter $y$ of \eref{eq:20110415-02}. Denoting a fixed-point solution of the belief-propagation equation (a macroscopic state) as $\alpha$ and its associated BP free energy as $F_{B P}^{(\alpha)}$, the grand partition function $\Xi$ can be re-written as
\begin{equation}
\label{eq:20110731-01}
\Xi=\sum\limits_{\alpha} \exp\Bigl(- y F_{BP}^{(\alpha)} \Bigr)
= \int {\rm d} f \exp\Bigl[N (\Sigma(f) - y f) \Bigr] ,
\end{equation}
where $\exp\Bigl(N \Sigma(f)\Bigr)$ is the total number of macroscopic states
with a given BP free energy  $F_{B P}=N f$ (the quantity $f$ is called the free energy density). The function $\Sigma(f)$ is called the complexity in the spin-glass literature (it is the entropy density at the level of macroscopic states). For $N\gg 1$, the integration in \eref{eq:20110731-01} are dominated by the value of $f=\overline{f}$ which satisfies $\frac{{\rm d}\Sigma(f)}{{\rm d} f}|_{f=\overline{f}}=y$. The value $\overline{f}$ is the mean BP free energy density at a given value of $y$. If we neglect the loop correction to the grand free energy in (\ref{eq:gfe}),
then
\begin{equation}
N \overline{f} \approx \frac{\partial[ y G_{SP}(y; \beta)]}{\partial y}
=\sum\limits_{i} \overline{f}_i -
\sum\limits_{(i,j)} \overline{f}_{(i,j)} ,
\end{equation}
where $\overline{f}_{i}$ and $\overline{f}_{(i,j)}$ are, respectively, the mean free energy contribution of a vertex $i$ and an edge $(i,j)$, with the expression
\begin{eqnarray}
 \overline{f}_i &=&
 \frac{\prod_{j\in \partial i} \int {\rm D} q_{j\rightarrow i}
P_{j\rightarrow i}(q_{j\rightarrow i}) f_i e^{-y f_i}
 }
 {
 \prod_{j\in \partial i} \int {\rm D} q_{j\rightarrow i}
P_{j\rightarrow i}(q_{j\rightarrow i}) e^{-y f_i}} , \\
\overline{f}_{(i,j)} &=&
\frac{
\int \int {\rm D} q_{i\rightarrow j} {\rm D} q_{j\rightarrow i}
P_{i\rightarrow j}(q_{i\rightarrow j})
P_{j\rightarrow i}(q_{j\rightarrow i}) f_{(i,j)}
e^{-y f_{(i, j)}}}{
\int \int {\rm D} q_{i\rightarrow j} {\rm D} q_{j\rightarrow i}
P_{i\rightarrow j}(q_{i\rightarrow j})
P_{j\rightarrow i}(q_{j\rightarrow i})
e^{-y f_{(i, j)}}} .
\end{eqnarray}
The value of the complexity $\Sigma$ is expressed as
\begin{equation}
\Sigma = y \Bigl[ \overline{f} - \frac{1}{N}G_{SP}(y; \beta)\Bigr] .
\end{equation}
The smallest mean free energy density $\overline{f}$ corresponds to the value of $y$ which makes the complexity be zero, $\Sigma=0$. Another special value of $y$ is $y=\beta$. If the complexity calculated at $y=\beta$ is positive, the corresponding mean free energy density value $\overline{f}$ is the typical value of BP free energy densities of the macroscopic states sampled at inverse temperature $\beta$ \cite{Mezard-Parisi-2001}.

\section{Conclusion and discussion}
\label{sec:conclude}

The main results of this paper are the free energy expression
(\ref{eq:ffull}) and the grand free energy expression
(\ref{eq:gfe}), and the corresponding belief-propagation equation
(\ref{eq:BPfix}) and survey-propagation equation (\ref{eq:SPfix}).
From the viewpoint of partition function loop expansion, the belief-propagation
and survey-propagation equation are, respectively, conditions needed to ensure that
subgraphs with dangling edges have zero correction contributions to the
free energy and the grand free energy.

This work helps to place the mean-field RSB theory of
spin glasses on a firmer mathematical ground. There are many unsolved problems ahead. For example,
the relationship free energy functionals $G_{BP}$ and $G_{SP}$ and the free energy landscape of the system; the
link between the defined grand partition function $\Xi$ and the original partition function $Z$; the issue of
sampling microscopic configurations $\{x_{i j}\}$ giving a
fixed-point solution $\{q_{j\rightarrow i}(x_{i j})\}$ of the belief-propagation equation; and son on.

The discussion in Sec.~\ref{sec:rsb} can be readily extended to the case that
the survey-propagation equation (\ref{eq:SPfix}) has multiple fixed-point solutions.
As a result, the mean-field second-step RSB free energy and its loop correction expression will be derived, as well as the corresponding message-passing equation.
This expansion hierarchy can be continued to produce the mean-field
results and the corresponding
loop correction expressions and message-passing equations at even higher-levels of replica-symmetry-breaking.

For statistical models defined on a factor graph with partition functions expressed
in the form of (\ref{eq:20110417-01}), the method of this paper can also be directly
applied without the need of first turning the partition function into the form of
(\ref{eq:partitionf}).

The present paper also points to some other important open issues. One question is: How to express the mean value of a local observable in terms of a finite loop series?
Examples of local observables are the state variable $x_{i j}$ on an
edge $(i,j)$ of the system, the correlation between two
edge variables $x_{i j}$ and $x_{k l}$, the energy of a single interaction,
and so on. Loop series expressions for these local observations should be
very useful in improving the predictions of the mean-field cavity theory. For a graphical model with many short loops, it is desirable to represent the system as a collection of many basic clusters in the framework of Kikuchi's cluster variation method (for a review, see \cite{Pelizzola-2005}). These basic clusters are connected to each other by joint clusters \cite{Morita-1981}. The joint clusters can be regarded as generalized edges.
The present partition function loop expansion method probably is also applicable to these more complex graphical
systems.

\ack

This work was finished while the authors were participating the
``Interdisciplinary Applications of Statistical Physics and Complex Networks''
program of KITPC (March, 2011).
JQX thanks Prof. Xiang-Mao Ding for encouragements and support.
This work was partially supported by the NSFC Grant 10834014,
the 973-Program Grant 2007CB935903, and the PKIP Grant KJCX2.YW.W10 of CAS.

\appendix
\section{Discrete models on a one-dimensional ring}
\setcounter{section}{1}
\label{sec:ring}

As a simple application of the graph expansion method discussed in the main text,
we calculate the loop correction contribution for a model defined on a one-dimensional ring with $N$ vertices and $N$ edges. We assume that the edge state $x_{i, i+1}$ between
 two vertices $i$ and $(i+1)$ can take
$Q$ different discrete values, $x_{i, i+1} \in \{1, 2, \ldots, Q\}$. The energy of the
ring is
\begin{equation}
E= - J \delta_{x_{1,2}}^{x_{N,1}}-
\sum\limits_{i=2}^{N-1} J \delta_{x_{i,i+1}}^{x_{i-1,i}} - J\delta_{x_{N-1,N}}^{x_{N,1}},
\end{equation}
where $J$ is a coupling constant and $\delta_x^y$ is the Kronecker symbol. The prior
distribution $\rho_0(x_{i,i+1})$ is uniform over the $Q$ states.

For this simple system, the fixed point of the
belief-propagation equation (\ref{eq:BPfix}) is
$q_{i\rightarrow (i+1)}(x)=1/Q$ and $q_{(i+1)\rightarrow i}(x)=1/Q$ for
$x\in \{1,2,\ldots, Q\}$. Then we have $\Delta_{i,i+1}(x_{i,i+1},x_{i+1,i})
= Q \delta_{x_{i,i+1}}^{x_{i+1,i}}-1$, and by a straightforward summation along the ring, the loop correction expression (\ref{eq:20110801-01}) is simplified as
\begin{equation}
L_g=(Q-1) \Bigl[\frac{e^{\beta J}-1}{Q-1+e^{\beta J}}\Bigr]^{N}.
\end{equation}
We notice that for $N$ being even, $L_g \geq 0$ and therefore the loop correction to the
free energy [see (\ref{eq:ffull})] is negative (the Bethe-Peierls free energy $F_{BP}$
is higher than the true free energy). However, if $N$ is odd and $J<0$, then $L_g < 0$
and the loop correction to the free energy becomes positive ($F_{BP}$ is lower than the
true free energy). This different behaviour is related to the fact that, the one-dimensional ring with an odd number of interactions is frustrated when $J<0$.
This simple example also shows that the loop correction $L_g$ decays exponentially with
loop length.

\section*{References}


\end{document}